
\normalbaselineskip 12pt
\hsize=16.5truecm
\vsize=24truecm
\parskip 0pt

\input phyzzx
\input epsf
\epsfverbosetrue
\Pubnum={\vbox{\hbox{CERN-TH.6991/93}\hbox{FTUAM. 93/28}}}
\pubnum={\vbox{\hbox{CERN-TH.6991/93}\hbox{FTUAM. 93/28}}}
\date={July, 1993}
\pubtype={}
\titlepage

\def\caption#1{\vskip 0.1in\centerline{\vbox{\hsize 2in\noindent
     \tenpoint\baselineskip=14pt\strut #1\strut}}}

\title{SOME  GLOBAL ASPECTS OF DUALITY IN STRING THEORY}
\vskip 1.0cm
\author{E. Alvarez\footnote{*}{Permanent address
Departamento de F\'{\i}sica Te\'orica,
Universidad Aut\'onoma de Madrid,
Canto Blanco, Madrid, Spain}, L. Alvarez-Gaum\'e ,
J.L.F. Barb\'on$^*$\foot{Address
after Sept. 1st, Physics Department,
Princeton University, Princeton, NJ 08544, USA},
Y. Lozano$^*$}
\address{Theory Division, CERN\break
 CH-1211 Geneva 23, Switzerland}

\abstract{We explore some of the global aspects of duality
transformations in String Theory and Field Theory.  We
analyze
in some detail the equivalence of dual models
corresponding
to different topologies at the level of the partition
function
and in terms of the operator correspondence for abelian
duality.  We analyze the behavior of the cosmological
constant under these transformations.
We also explore several examples of non-abelian duality
where the
classical background interpretation can be maintained
for the original and
the dual
theory.  In particular, we construct a non-abelian dual
of $SL(2,R)$ which turns out to be a three-dimensional
black hole.}

\endpage
\chapter{INTRODUCTION}
Duality symmetry plays an important r\^ole in
Statistical Mechanics
\REF\dizyk{J.M. Drouffe
and C. Itzykson: {\it Quantum Field Theory and Statistical
Mechanics},
Cambridge U. Press, 1990.}(for a review and references to
the
literature see for instance [\dizyk]), in particular in
the
analysis of the phase diagram of spin systems.  It can also be
understood as a way to show the equivalence between two
apparently different theories. On a
lattice system described by a Hamiltonian $H(g_i)$
with coupling constants $g_i$
 the duality transformation
produces a new Hamiltonian  $H^*(g^*_i)$ with coupling
constants $g^*_i$ on the dual lattice.  In
this way one can often relate the strong coupling regime
of $H(g)$ with the weak coupling regime of $H^*(g^*)$.
In String Theory and Two-Dimensional Conformal Field Theory
duality is an important tool to show the equivalence of
different geometries and/or topologies and in determining
some of the genuinely stringy implications on the structure
of the low energy Quantum Field Theory limit.
Duality symmetry was first described on the
context of toroidal compactifications \REF\bgs {L. Brink,
M.B. Green, J.H. Schwarz: Nucl. Phys. {\bf B198}
 (1982), 474.} \REF\tordual {K. Kikkawa and M. Yamasaki:
Phys. Lett. {\bf B149} (1984), 357 ; N. Sakai
and I. Senda: Prog. Theor. Phys. {\bf 75} (1984), 692.}
[\bgs, \tordual]. For the simplest case of a single
compactified dimension of radius $R$, the entire physics
of the interacting theory is left unchanged under the
    replacement $R \rightarrow {\alpha}^{'} /R $
provided one also transforms the dilaton field $\phi
\rightarrow \phi - {\rm log}(R/\sqrt{{\alpha}^{'}})$
\REF\eamar {E. Alvarez and M.A.R. Osorio: Phys. Rev. {\bf D40}
 (1989), 1150.} [\eamar]. This simple case can
be generalized to  arbitrary toroidal compactifications
described by constant metric $g_{ij}$ and antisymmetric
tensor $b_{ij}$ \REF\nsw {K. Narain, H. Sarmadi and E. Witten:
Nucl. Phys. {\bf B279} (1987), 369.} [\nsw]. The
generalization of the duality symmetry in this case becomes
$(g+b) \rightarrow (g+b)^{-1}$ and $\phi \rightarrow \phi
-{1\over 2}{\rm log \, det}(g+b)$. In fact this transformation
is an element of an infinite order discrete symmetry group
for d-dimensional toroidal compactifications $O(d,d; Z)$
\REF\torodd {V. P. Nair, A. Shapere, A. Strominger and
F. Wilczek: Nucl. Phys. {\bf B287} (1987), 402;
A. Shapere and F. Wilczek: Nucl. Phys. {\bf B320}
(1989), 609; A. Giveon, E. Rabinovici and G. Veneziano: Nucl.
Phys. {\bf B322} (1989), 167.}[\torodd]. The
symmetry was later extended
 to the case
 of non-flat
conformal backgrounds
\REF\buscher{ T. H. Buscher: Phys. Lett. {\bf B194}
 (1987), 51 ; Phys. Lett. {\bf B201}
 (1988), 466.} in [\buscher].
  In Buscher's construction one starts
with a  manifold $M$ with metric
$g_{ij}, i,j=0,\ldots d-1$, antisymmetric
tensor $b_{ij}$ and  dilaton field $\Phi(x_i)$.
One requires the metric to admit at least one
continuous isometry which leaves invariant the
$\sigma$-model action constructed out of $(g, b, \phi)$.
Choosing an adapted coordinate system $(x^0, x^{\alpha}) =
(\theta, x^{\alpha}), \alpha = 1, \ldots d-1 $
where the isometry acts by translations of $\theta$, the
change of $g, b, \phi$ is given by
\def\dg{{\tilde g}}
\def\db{{\tilde b}}
$$
\eqalign{
\dg_{00}&=1/g_{00},\qquad
         \dg_{0\alpha}=b_{0\alpha}/g_{00},\qquad
          \dg_{\alpha\beta} = g_{\alpha\beta} -
{g_{0\alpha}g_{0\beta} - b_{0\alpha} b_{0\beta}\over g_{00}}\cr
          \db_{0\alpha} &= {g_{0\alpha} \over g_{00}},\qquad
        \db_{\alpha\beta}=b_{0\beta}-{g_{0\alpha}b_{0\beta}
         -g_{0\beta}b_{0\alpha}\over g_{00}},\cr
        {\tilde \phi}&=\phi-{\rm log}g_{00}.\cr}
\eqn\busdual
$$
The result is that for any continuous isometry of the metric
which is a symmetry of the action one obtains the equivalence
of two apparently very different non-linear $\sigma$-models.
If n is the maximal number of commuting isometries, one obtains
a duality group of the form $O(n,n;Z)$ \REF\givroc {A. Giveon, M.
Ro\v cek: Nucl. Phys. {\bf B380} (1992), 128.}
[\givroc] (a rather thorough review of recent results
in duality can be found in
\REF\kirit{E. Kiritsis, {\it Exact Duality Symmetry
in CFT and String Theory}, CERN-TH-6797/93.}
[\kirit]).
 Duality symmetries are useful in determining important
properties of the low-energy effective action, in particular in
questions related to supersymmetry breaking and in the removal of
flat directions from the potential \REF\susydual {A. Font,
L.E. Iba\~nez, D. L\"ust and F. Quevedo: Phys. Lett. {\bf B245}
 (1990), 401; S.Ferrara, N. Magnoli,
T.R. Taylor and G. Veneziano: Phys. Lett. {\bf B245}
(1990), 409; H.P. Nilles and M. Olechowski:
Phys. Lett. {\bf B248} (1990), 268;
P. Binetruy and M.K. Gaillard: Phys. Lett. {\bf B253}
(1991), 119.} [\susydual]. Although the
transformation \busdual\ was originally obtained using
 a method apparently
not compatible with
general covariance, it is not difficult to modify the
construction to eliminate this drawback.
A particularly useful
interpretation of \busdual\ is in terms of the gauging
of the isometry symmetry
\REF\rocver {M. Ro\v cek and E. Verlinde: Nucl. Phys.
{\bf B373} (1992), 630.} [\rocver]
with an auxiliary gauge field whose field strengh is
required to vanish,
although it will in general have non-trivial monodromy
in non-spherical topologies. The analysis
in [\rocver] relates the two dual $d$-dimensional
$\sigma$-models
with an auxiliary $d+1$-dimensional
$\sigma$-model which contains a left- and a
right-handed
chiral current associated to the original isometry.
The original model and its dual are then obtained by
gauging respectively the vector and axial vector
combinations of chiral currents on the
auxiliary theory. This presentation also clarified
the relation between the conformal properties
of the three models involved.

A further generalization of the general procedure in
[\rocver] was proposed in \REF\queossa {X.C. de la
Ossa and F. Quevedo: Nucl. Phys. {\bf B403} (1993), 377.}
[\queossa] in the case that the manifold
$(M,g)$ has non-abelian isometries. This form of
 ``duality'' contains many striking differences with
ordinary or abelian duality, and it is one of our
purposes to explore some of its properties.

The outline of this paper is the following.  In
section two we present the general form
of the duality transformation for compact
abelian isometries.  We use a manifestly
covariant formalism and therefore we can control
the global topological properties of the
dual model. In particular we are interested
in the case where the original manifold has
a trivial fundamental group.  Naively one
would not expect to find any topology change
in this case because the original model does
not contain any winding states, and toroidal
duality can be expressed as the exchange
of winding and momentum states.  We will see
that generically the duality transformation
changes the topology and/or the geometry
of the manifold in a way
largely independent of its fundamental group.
In section three we study in detail a particular
example (the $\sigma$-model on the three-sphere)
as an illustrative case of a manifold without
fundamental group whose topology changes to
$S^2\times S^1$.  We perform the duality
transformation patch by patch and we explicitly
reconstruct the topology of the dual manifold.
Section
four contains the explicit study of abelian
duality in Wess-Zumino-Witten models where
one can explicitly control the global topology
rather easily.  Section five presents the
explicit map between operators of the original
and dual theories (order/disorder mapping)
for arbitrary curved backgrounds.
In section six we make some remarks concerning
the behavior of the cosmological constant
under duality transformation and explore
briefly some of its consequences.
Finally
in section seven we come to the study of
non-abelian duality.  We present a series
of interesting examples based on manifolds
with non-abelian isometry groups acting
without fixed-points.  The simplest
example is provided by the $SU(2)$-non-linear
$\sigma$-model and as isometry group
we take the left action of $SU(2)$ on
itself.  The dual model does not contain
singularities, but it is non-compact
(it is the Lie algebra of $SU(2)$
with a curved metric). Next we study
the same procedure for the $SL(2,{R})$
$\sigma$-model.  $SL(2,{R})$
is geometrically anti-DeSitter space,
and under non-abelian duality we obtain
a three-dimensional anti-DeSitter
black hole.  In this case, although
the isometries act without fixed points,
the singularity appears as a consequence
of the non-compactness of the isometry
group.  After some brief survey of other examples,
we study some of the still unresolved
questions in trying to construct the
non-abelian duality transformation for
higher genus Riemann surfaces, and the
difficulties with the explicit
construction of the operator mapping
between the original model and its dual.

\chapter{GENERALLY COVARIANT DUALITY TRANSFORMATION}
In \busdual\ we saw that the original construction
of the duality transformation required a particular
choice of coordinates. Since we are interested in the
global topological properties of the dual theory
$({\tilde M}, \dg, \db)$, we would like to begin by
writing the form of \busdual\ in a general covariant
form.  This is all the more interesting in those cases
we will consider later where ${\pi}_1 (M) =0$. From
a String Theory point of view there are only momentum
modes and no winding modes, and thus we wish to
clarify first of all what is the global topology of
$\tilde M$ and the form of the operator mapping in this
case. The way to proceed following the ideas in [\rocver]
is to gauge the isometry. With a general $b_{ij}$ field
(Wess-Zumino-Witten term) we need to use the Noether
procedure as made explicit in \REF\isogauge {I. Jack,
D.R.T. Jones, N. Mohammedi and H. Osborn: Nucl. Phys.
{\bf B332} (1990), 359; C. Hull and B.
Spence: Phys. Lett. {\bf B232} (1989), 204.}
[\isogauge]. After gauging the isometry we impose the
constraint that the gauge field should be flat. In this
formalism going to adapted coordinates is a question of
gauge fixing. However issues related to the resulting
global topology and geometry can be addressed without
recurring to a particular gauge choice, and we can
always keep explicit general covariance.

 Ignoring the dilaton, we consider the $\sigma$-model
$$
\eqalign{
S &= {1 \over 2\pi} \int d^2\sigma
(g_{ij} + b_{ij}) \partial x^i
{\overline \partial} x^j =\cr
\, &= {1 \over  8\pi}\int g_{ij} \partial_{\mu} x^i
\partial^{\mu} x^j + {i \over 8\pi} \int b_{ij} dx^i
\wedge dx^j ,\cr}
\eqn\eqnoriginal
$$
the second term on the right-hand side is the Wess-Zumino
term. If the world-sheet is a genus $g$ Riemann surface
$\Sigma_{g}$, the Wess-Zumino term can be described in
terms of an element of $H^3(M, {R})$. If $\Sigma_{g}^0 =$
interior of $\Sigma_g$, and $H$ is the selected
3-form in $H^3(M,{R})$, the Wess-Zumino-Witten term in the
action is
$$
\Gamma = {i\over 8\pi} \int_{\Sigma_{g}^0 , \partial
\Sigma_{g}^0 = \Sigma_g} H
\eqn\wz
$$
We assume $H$ can be chosen in such a way that the
ambiguity in $\Gamma$ due to different choices of
$\Sigma_{g}^0$ is equal to $2\pi i$(integer), and the
quantum theory is thus well-defined. Let $k^i$ be a
Killing vector for the metric $g$:
$$
{\cal L}_k g_{ij} = \nabla_i k_j + \nabla_j k_i =0
\eqn\killing
$$
Invariance of $S$ requires also
$$
\delta_k \Gamma = {i\over 8\pi} \int_{\Sigma_{g}^0}
\epsilon {\cal L}_k H = {i\over 8\pi} \int_{\Sigma_{g}^0}
\epsilon (d i_k + i_k d) H ={i\over 8\pi}\epsilon
\int_{\Sigma_{g}} i_k H=0,
\eqn\varwz
$$
because $dH=0$, and $(i_k H)_{ij} = k^l H_{lij}$, hence
\varwz\ vanishes if $i_k H = -dv$, for some 1-form $v$. In
terms of $b$, $(H=db$ locally) this implies:
$$
{\cal L}_k b = d\omega, \,\,\,\,\,\,\,\, \omega =
i_k b - v.
\eqn\tururu
$$
The associated conservation law is:
$$
\partial{\bar J}_k + {\bar \partial} J_k=0
\eqn\lawcon
$$
$$
\eqalign{
J_k &= (k - i_k b + \omega)_i \partial x^i =
(k - v)_i \partial x^i \equiv (k - v)\cdot \partial x \cr
{\bar J}_k &= (k + i_k b - \omega)_i
{\bar \partial} x^i =
(k + v)_i {\bar \partial} x^i \equiv (k + v)\cdot
{\bar \partial} x \cr}
\eqn\jotas
$$
 If we wish to gauge the isometry we introduce gauge fields
$A, {\bar A}$, with $ \delta_{\epsilon}A = -\partial \epsilon
\,\, , \, \, \delta_{\epsilon}{\bar A} = -{\bar \partial}
\epsilon$, and $ \delta x^i = \epsilon k^i (x)$ now with
$\epsilon$ a function on the world-sheet. The original action
will vary into $ {1\over 2\pi} \int (J_k {\bar \partial}
\epsilon + {\bar J}_k \partial \epsilon)$ which can be
cancelled by adding an extra term to the action of the form
$\int ( A {\bar J}_k + {\bar A} J_k)/2\pi$. However
the new term still varies under the gauge isometry due to the
variation of $ J_k\,,\,\, {\bar J}_k$. By adding
one new term to the action, $\int k^2 A{\bar A}/2\pi $,
the total variation becomes
$$
\delta ( {\it L} + {\it L}^{'} + {\it L}^{''}) =
A{\bar \partial}(\epsilon k \cdot v) - {\bar A}
\partial (\epsilon k \cdot v),
\eqn\anomala
$$
where $\it L$ is given by \eqnoriginal\ and:
$$
\eqalign{{\it L}^{'}&={1\over 2\pi}(A {\bar J}_k+
{\bar A}J_k) \cr
{\it L}^{''}&={1\over 2\pi}k^2 A {\bar A}\cr}
\eqn\alphas
$$
The anomalous variation \anomala\ cannot be cancelled in general
unless we include extra fields into the action. The simplest
choice is a real scalar field $\chi$, contributing to
the Lagrangian a term:
$$
{\it L}_{\chi}={1\over 2\pi}(A{\bar\partial}\chi-{\bar A}
\partial\chi),
\eqn\chachi
$$
and
$$
{\delta}_{\epsilon}\chi=-\epsilon k\cdot v
\eqn\varia
$$
The complete action and transformation rules are:
$$
S_{d+1}={1\over 2\pi}\int[(g_{ij}+b_{ij})\partial
x^i{\bar\partial}x^j+(J_k-\partial\chi){\bar A}
+({\bar J}_k+{\bar\partial}\chi) A + k^2 A
{\bar A}] d^2 \sigma
\eqn\dualac
$$
$$
\eqalign{\delta_{\epsilon} x^i &= \epsilon k^{i}(x) \qquad
\delta_{\epsilon} \chi = -\epsilon k \cdot v \cr
\delta_{\epsilon} A &= -\partial \epsilon \qquad
\delta_{\epsilon} {\bar A} = -{\bar \partial} \epsilon \cr}
\eqn\gaugetrans
$$
For a genus $g$ world-sheet $\Sigma_g$ and compact isometry
orbits we may have large gauge transformations. We consider
multivalued gauge functions:
$$
\int_{\gamma} d \epsilon = 2\pi n(\gamma) \qquad
n(\gamma) \in {Z},
\eqn\epsperiod
$$
where $\gamma$ is a non-trivial homology cycle in $\Sigma_g$.
Since we are dealing with abelian isometries it suffices to
consider only the toroidal case $g=1$. The variation of
$S_{d+1}$ \dualac\ is:
$$
\eqalign{\delta S_{d+1} &= {1\over 2\pi} \int \left( \partial
\chi {\bar \partial}\epsilon - \partial \epsilon {\bar
\partial} \chi\right) = {i\over 4\pi} \int_{T} d\chi \wedge
d\epsilon \cr
&= {i\over 4\pi} \left( \oint_{a} d\chi \oint_{b} d\epsilon
- \oint_{a} d\epsilon \oint_{b} d\chi \right) \cr}
\eqn\Riemannid
$$
where $a$ and $b$ are the two generators of the homology
group of the torus T. Since $\epsilon$ is multivalued by
$2\pi {Z}$, we learn from \Riemannid\ that $\chi$ is
multivalued
by $4\pi {Z}$.
$$
\oint_{\gamma} d\chi = 4\pi m(\gamma)
\qquad m(\gamma) \in {Z}.
\eqn\chiperiod
$$
For a non-compact isometry $\delta S_{d+1} =0$ and $d\chi$ may
in general have real periods. Thus gauge invariance determines
the explicit multivaluedness and the periods of $\chi$.

 In duality arguments the first step consists of exhibiting the
equivalence of $S_{d+1}$ \dualac\ with the original model
\eqnoriginal\ . This in principle follows on spherical topologies
by eliminating $\chi$ from $S_{d+1}$. For other topologies we
have to be careful with the possible monodromy contributions
since $\chi$ is multivalued. If we consider the two terms in
$S_{d+1}$ containing $\chi$ and $A, {\bar A}$, we can decompose
$d\chi$ into an exact part $d \chi_0$ ($\chi_0$ a single valued
function) plus a harmonic piece $d\chi_h$ which has non-trivial
periods. The term $d\chi_0 \wedge A$ can be integrated by parts,
and upon eliminating $\chi_0$ it imposes the constraint that $A$
should be flat. However, although $A$ may be flat, it can have
non-trivial holonomy about the $a, b$ cycles. The harmonic part
of $\chi$ then enforces the triviality of the holonomy of $A$,
and this then implies that A is pure gauge and that indeed
\dualac\ is equivalent to \eqnoriginal\ . The second step, which
produces the dual action, implies the elimination of the gauge
field $A$. This is possible because $A$ appears only
quadratically
in the action. However, since in principle the Killing vector
could have fixed points (points where $k^2 =0$),
it is perhaps
better to think of $S_{d+1}$ \dualac\ as the dual action to
\eqnoriginal.

 By construction $S_{d+1}$ is manifestly general covariant,
and therefore we have a clear idea of the $d$-dimensional
geometrical interpretation of the model. We are working
on the quotient of $M$ by the orbits of the isometry group
times the $\chi$-space. For a compact isometry $\chi$ lives
on a circle and therefore generically one would expect to
end up in the product manifold (or orbifold)
${\tilde M}=``(M/S^1)\times S^{1}_{\chi}"$.
Generically we
expect topology change as a consequence of duality. However
the more delicate issue is whether the dual manifold
$\tilde M$ is indeed a product or a twisted product (non-
trivial bundle, etc.). It is  also useful to notice that in
the previous arguments
 the structure of $\pi_{1}(M)$ played
no r\^ole. In toroidal compactifications
$\pi_1 (T^d) = {Z}^d$,
and together with ordinary momentum states we have winding
states describing the way the string wraps around in the
compactified dimensions. Duality is described as the symmetry
exchanging momentum and winding states. For other
manifolds, in particular those where $\pi_{1}(M)=0$
and yet ${\tilde M}$ is not diffeomorphic to $M$, we have
to reconsider the mapping between the operators of the
two theories.

If we are interested in the explicit form of the dual
$\sigma$-model we have to eliminate the gauge field $A$.
For Killing vectors with fixed points ($k^2 =0$) this
generates singular manifolds. Thus gauge fixing \dualac\
in this case produces a singular space whose background
interpretation is doubtful whereas \dualac\ itself is
perfectly well-defined. The simplest way to gauge fix
\dualac\ is to choose coordinates adapted to the Killing
field $k^{i}$: $(\theta, x^{\alpha})$, $k^{i} \partial
/ \partial x^i = \partial / \partial \theta $. This is a
representation that depends on the system of coordinate
patches used, and the global issues appear as Gribov
problems. Using adapted coordinates, and eliminating
the gauge field, we obtain after some manipulation the
following form for $\dg, \db$:

$$
\eqalign{
\dg_{00}&={1\over k^2}  \cr
         \dg_{0\alpha}&={v_{\alpha} \over k^2} \cr
\db_{0\alpha} &= {k_{\alpha} \over k^2} \cr
          \dg_{\alpha\beta} &= g_{\alpha\beta} -
{k_{\alpha}k_{\beta} - v_{\alpha} v_{\beta}\over k^2}\cr
        \db_{\alpha\beta}&=b_{\alpha\beta}-{k_{\alpha}v_{\beta}
         -k_{\beta}v_{\alpha}\over k^2}\cr}
\eqn\rokdual
$$

In local adapted coordinates we recover Buscher's
transformation. As pointed out before, the global structure
is captured directly from $S_{d+1}$ without the need to
integrate out the gauge field.

A different way to proceed is to cover the target manifold
$M$ with coordinate patches, do patchwise duality
transformations, and then try to reconstruct $\tilde M$ out
of the ``dual" patches. We will use this procedure in a simple
example in the next section for illustrative purposes.

\chapter{SOME SIMPLE EXAMPLES IN NON-CONFORMAL THEORIES}

In the previous section we analyzed
the duality transformation with respect
to an abelian symmetry in a manifestly
covariant framework to understand the global
topology of the dual manifold.  It is also
instructive to understand the same results
explicitly in some examples, especially
in those with $\pi_1(M)=0$.  Since we want
to maintain the background interpretation
for the $\sigma$-model, we look for geometries
with fixed-point free isometries and vanishing
fundamental group.  The simplest example is
to take $M=SU(2)=S^3$ with the round metric.
Locally $S^3$ is certainly equivalent to
$S^2\times S^1$, but not globally.  The left action
of $SU(2)$ on the round metric is free from
fixed points.  We may represent $S^3$ as
a submanifold of ${C}^2$, with coordinates
$(z_0,z_1)$, satisfying ${\overline z_0}z_0
+{\overline z_1}z_1=1$.  It is also useful
to represent $S^3$ as a Hopf fibering over
$S^2$, $p:S^3 \rightarrow S^2$. When
$z_0\ne 0$, $p(z_0,z_1)=z=z_1/z_0$, and
when $z_1\ne 0$, $p(z_0,z_1)=1/z=z_0/z_1$.
If $H_{\pm}$ represent the two hemispheres
of the two-sphere, the local equivalence
of $S^2\times S^1$ with $S^3$ is given by
$$
\eqalign{H_+:(z,u_+)\in S^2\times S^1&\rightarrow
({u_+\over \sqrt{1+|z|^2}},
{zu_+\over \sqrt{1+|z|^2}})\cr
H_-:(1/z,u_-)\in S^2\times S^1&\rightarrow
({|z|u_-\over \sqrt{1+|z|^2}},
{|z|u_-/z\over \sqrt{1+|z|^2}}).\cr}
\eqn\hopffib
$$
The transition function in the equator of
the two-sphere is
$$
u_+={|z|\over z} u_-=e^{i\phi}u_-,
\eqn\hopftran
$$
which defines the necessary twist in the
fiber in order to obtain the three-sphere
as the global space.  We can parametrize
the geometry in terms of Euler angles
$$
\eqalign{z_0&=e^{i(\chi +\phi)/2}\cos\theta/2\cr
z_1&=e^{i(\chi -\phi)/2}\sin\theta/2\cr},
\eqn\eulerang
$$
where $0\le\theta<\pi, 0\le\phi<2\pi,
0\le\chi<4\pi$, corresponding to a parametrization
of $SU(2)$ as follows
$$
g=e^{i\chi \sigma_3/2}
e^{i\theta \sigma_1/2}
e^{i\phi \sigma_3/2}.
\eqn\eulerpauli
$$
In terms of these coordinates the left-invariant
one-forms $g^{-1}dg=i\sigma_a \omega_{aL}/2$
are given by
$$
\eqalign{\omega_{1L}&=\sin\theta\sin\phi\, d\chi
+\cos\phi\, d\theta\cr
\omega_{2L}&=\sin\phi\, d\theta-
\sin\theta\cos\phi\, d\chi\cr
\omega_{3L}&=d\phi+\cos\theta\, d\chi,\cr}
\eqn\leftforms
$$
and the round metric is simply
$$
\eqalign{ds^2&=\delta_{ab}
\omega_{L}^a\omega_{L}^b\cr
&=d\theta^2+d\phi^2+2\cos\theta d\phi d\chi
+d\chi^2 .\cr}
\eqn\roundmetric
$$
The isometry we will be interested in
corresponds in complex coordinates to
the change of the common phase of
$z_0$ and $z_1$.  In terms of Euler angles
this is $\chi\rightarrow\chi +{\rm const.}$.
Since the Euler angles are adapted coordinates
to this isometry we can apply Buscher's formulae
(1.1) to obtain the dual metric
$$
d{\tilde s}^2=d\theta^2 +d{\tilde \chi}^2 +
\sin^2\theta d\phi^2,
\eqn\sthreeb
$$
and
$$
{\tilde b}=2\cos\theta d{\tilde \chi}\wedge d\phi.
\eqn\sthreeb
$$
this dual metric looks certainly like
$S^2\times S^1$, but at this point it is a bit difficult
to discriminate between $S^2\times S^1$ and $S^3$
because Euler angles (like polar coordinates)
are not globally valid, and we should beware
of global topological conclusions obtained
in a particular coordinate system.  We
can be more careful and use a description
of $S^3$ in terms of two coordinate
patches using stereographic projection.
In ${R}^3$ we embed $S^3$ by the condition
$(\xi^0)^2+ (\xi^i)^2=1, i=1,2,3$.  The
stereographic projection is explicitly
given by
$$
x_i^{\pm}={\xi_i\over 1\mp\xi_0},
\eqn\stereocoord
$$
with transition functions
$$
x_i^+=x_i^-/x_-^2 \qquad x_-^2 x_+^2=1.
\eqn\transfn
$$
The induced metric on $S^3$ is the round metric.  The
Killing vector $k=\partial /\partial\chi$ in the two
hemispheres takes the form:
$$
k^i_{\pm}=(\pm x_1x_3-x_2,x_1\pm x_2x_3,\pm (1+x_3^2-
x_1^2-x_2^2)/2),\qquad k^2_{\pm}=1.
\eqn\killingstereo
$$
Applying Buscher's formulae in both patches
we obtain the dual metrics:
$$
d{\tilde s}^2_{\pm}=d{\tilde\chi}
_{\pm}^2
+(g_{ij}-k^{\pm}_ik^{\pm}_j)
dx^i_{\pm}dx^j_{\pm}.
\eqn\dulametpm
$$
${\tilde\chi_{\pm}}$ are the Lagrange
multipliers in each patch.  We also generate
$b$-fields
$$
{\tilde b_{\pm}}=-k^{\pm}\wedge
d{\tilde\chi_{\pm}}
\qquad k^{\pm}=k^{\pm}_{i}dx^{i}.
\eqn\dualbfield
$$
We can patch now $d{\tilde s}^2_{\pm}$
by imposing ${\tilde\chi_+}={\tilde\chi_-}$.
We also know that there is a redundant degree
of freedom in the metric because of gauge
invariance.  Thus, in order to obtain a metric
depending on three coordinates we have to make
a gauge choice, for example by choosing
$x^1=0$.  However this is possible as long
as $k^1(x)\ne 0$.  Hence to fix the gauge
we have to first look at the fixed surfaces
of the three components of the Killing vector.
The set of zeroes of $k^1_{\pm}$ is the
surface $Z_1^{\pm}=\{x_2=\pm x_1x_3\}$. Similarly
$Z_2^{\pm}=\{x_1=\mp x_2x_3\}$, and
$Z_3^{\pm}=\{x_3^2=x_2^2+x_1^2-1\}$.  These
sets have vanishing intersections because
the Killing vector has no fixed points.  In
fixing the gauge we can choose $x_1=0$ only
in $H_+-Z_1^+$.  In $Z_1^+-(Z_1^+\cap Z_2^+)$
we can choose $x_2=0$, and finally
in the intersection between $Z_1^+$
and $Z_2^+$ we may choose $x_3=0$.  In this
form the gauge is properly fixed and we
indeed obtain $S^2\times S^1$.
Incidentally, the correct procedure to fix
the gauge solves a small conceptual puzzle.
We can cover $S^3$ with two simply connected
coordinate patches which overlap in a simply
connected region.  However this is not the
case for $S^2\times S^1$.  The extra patches needed
to obtain a good  cover of this manifold
follow from carefully fixing the gauge in
order to describe the dual manifold.

The example we have just considered can be
extended to an arbitrary compact Lie group
$G$.  For any abelian subgroup $H$ in $G$
we may consider its left action
$h\in H, g\rightarrow hg$.  For the
principal chiral model,
$$
S={1\over 8\pi}\int d^2\sigma
Trg^{-1}\partial_{\mu} g
g^{-1}\partial^{\mu} g.
\eqn\princhmod
$$
Once again we can follow the procedure
of gauging the subgroup $H$, introducing
the Lagrange multiplier etc. to obtain
the dual model.  We will spare the reader
the details, and simply mention the
result.  After appropriate gauge fixing
in this case (labelling correctly
the $H$-orbits in $G$) we obtain as expected
the dual manifold
$$
{\tilde G}=(G/H)\times \{{\rm Lagrange \,\,
multiplier \,\, manifold}\}.
\eqn\dualgroup
$$
We should also be careful about the
behavior of the measure in the path
integral under this decomposition.  However
the measure of the original path integral
is the Haar measure. It is known
\REF\helga{S. Helgasson:
{\it Groups and Geometric Analysis},
Academic Press, 1984.}
[\helga] that when
$|det\, Ad_G(h)|=|det\, Ad_H(h)|$
for all $h\in H$\foot{$Ad_G(h)$ and
$Ad_H(h)$ are respectively the adjoint
actions of $h\in H$ considered as an element
of $G$ and $H$.}
there exists a measure on the quotient
$G/H$ such that
$$
\int_Gf(g)dg=\int_{G/H}dg_H\int_H dh f(hg_H).
$$
In the example at the beginning of the
section $H=U(1), G=SU(2)$, and the above
condition is satisfied.  This completes
the (semiclassical) proof of the topology
change from $S^3$ to $S^2\times S^1$.  We will later
analyze the mapping between operators of the
two models, however we would like to make a few
remarks about the mapping between states.
In the three-sphere, a natural basis to expand
the states is in terms of the Wigner functions
(the natural basis of functions on $SU(2)$).
These functions are labelled by three parameters
$D^j_{m,m'}$, and we can label the vertex
operators in terms of them.  When we consider
the dual space $S^2\times S^1$, the states
on $S^2$ are naturally labelled in terms of
spherical harmonics $Y^l_m$.  Naively, the
states in $S^1$ would be labelled by two
quantum numbers $(n,n')$, (momentum and winding
respectively)
and therefore
it would seem that there is a redundancy
of states with respect to the spectrum on
the three sphere we started with.  The resolution
of this puzzle is found on the fact that the
equation of motion of the Lagrange multiplier
field $\chi$ is not given in terms of the
free Laplace equation on the world-sheet,
but it contains a contribution due to the
induced WZW term, which is the analogue of
the Dirac monopole connection. In solving
this equation, the number of modes with
finite action will be cut in half, and therefore
the count of states in the original model
and its dual agrees.

\chapter{ABELIAN DUALITY IN WZW MODELS}

A class of Conformal Field Theories (CFT)
where the global properties of duality
can be controlled in detail are the
Wess-Zumino-Witten (WZW) models
\REF\wzw{E. Witten: Comm. Math. Phys.
{\bf 92} (1984), 455.}
[\wzw].  On a genus $g$ Riemann surface
the action for a simply connected group is
\def\del{\partial}
\def\delb{\overline{\partial}}
\def\gdg{g^{-1}\del g}
\def\gdbg{g^{-1}\delb g}
$$
S_0[g]=-{k\over 2\pi}\int_{\Sigma_g}
Tr(\gdg \gdbg) +
{ik\over 12\pi}\int_{\Sigma_g^0}
Tr(\gdg)^3,
\eqn\ei
$$
where $\Sigma_g^0$ is a filled surface
whose boundary is the Riemann surface
$\Sigma_g$.  In these models the
duality manipulations can be
carried out without reference to any coordinate
system.  This is due to the Polyakov-Wiegmann
property (PW)
\REF\pwp{A.M. Polyakov and P.B. Wiegmann:
Phys. Lett. {\bf 131B} (1983), 121.}
[\pwp]
$$
S_0[g_1g_2]=S_0[g_1]+S_0[g_2]
-{k\over \pi}\int_{\Sigma_g} Tr(g_1^{-1}\del g_1
\delb g_2 g_2^{-1}).
\eqn\eii
$$
For genus higher than one and for twisted
fields this formula receives corrections,
however for abelian twists \eii\
is correct in any genus.  We consider
the following  left and
right action of an abelian subgroup
generated by $H$:
$$
h_L=e^{\cos\alpha \theta H}\qquad
h_R=e^{\sin\alpha \theta H}.
\eqn\eiii
$$
Using PW
$$
S_0[h_L g h_R^{-1}]=S_0[g]+
{k\over 2\pi}\int\eta_{\alpha} \del \theta
\delb\theta +{k\over 2\pi}\int
(J_{\alpha}\delb\theta-{\overline J_{\alpha}}
\del\theta ),
\eqn\eiv
$$
where
$$
\eqalign{\eta_{\alpha}&=\sin2\alpha\,
TrHgHg^{-1} -TrH^2\cr
J_{\alpha}&=2\sin\alpha\, TrH\gdg\cr
{\overline J_{\alpha}}&=2\cos\alpha\, TrH\delb
g g^{-1}.\cr}
$$
This formula represents the effect
of a gauge transformation on $S_0$ with
respect to an arbitrary mixing of left/right
abelian actions for some abelian subgroup
$H_{\alpha}$.  Special cases are $\alpha=0,\pi/2$
corresponding to pure left- or right-chiral
rotations and $\alpha=\pm\pi/4$, which
represent vector and axial transformations.
For non-compact $H_{\alpha}$ we take
$H_{\alpha}^{\dagger}=H_{\alpha},H_{\alpha}^2=1$
while for compact subgroups we take
$H_{\alpha}^{\dagger}=-H_{\alpha}, H_{\alpha}^2=-1$.
In
this case we restrict to $\tan\alpha=n_R/n_L$,
a rational number, to avoid ergodic actions.
Gauging the WZW Lagrangian \ei\ with respect
to the action \eiii\ we obtain the dual model
$$
S^{(\alpha)}=S_0(g)-{k\over 2\pi}
\int A({\overline J_{\alpha}}+t_2\delb \chi)
+{k\over 2\pi}\int {\overline A}
( J_{\alpha}+t_2\del\chi)
+{k\over 2\pi}\int\eta_{\alpha}
A{\overline A},
\eqn\ev
$$
where $t_2=TrH^2$, with the gauge invariance
$$
\eqalign{g &\rightarrow e^{\cos\alpha\epsilon H}
g e^{-\sin\alpha\epsilon H}\cr
A &\rightarrow A-d\epsilon\cr
\chi &\rightarrow \chi-\epsilon\cos2\alpha .
\cr}
\eqn\evi
$$
In the compact case large gauge transformations
correspond to
$$
\oint_{\gamma}d\epsilon=2\pi r_{\alpha} n(\gamma)
\qquad r_{\alpha}=\sqrt{n_L^2+n_R^2},
\eqn\evii
$$
where $\gamma$ is a non-trivial homology cycle
and $n(\gamma)$ is an integer.

If we integrate the Lagrange multiplier we
project into $A=da$, with $\oint_{\gamma}
da=2\pi r_{\alpha}n(\gamma)$, and using
the PW property together with the invariance
of the Haar measure we easily establish the
equivalence between \ei\ and \ev.  The explicit
form of the dual model follows from $A$-integration,
and yields
$$
{\tilde S}=S_0(g_c)+{k\over 2\pi}\int
{(J_c+t_2\del\chi)({\overline J}_c+t_2\delb\chi)
\over \eta_{\alpha}},
\eqn\eviii
$$
where $J_c=J_{\alpha}(g_c)$, $g_c\in
G/H_{\alpha}$.
The multivaluedness of $\chi$
is as follows
$$
\oint_{\gamma}d\chi={4\pi\over
r_{\alpha} k t_2}n(\chi)
\eqn\eix
$$
in the case with $H_{\alpha}$ compact,
($n(\chi)$ is an integer) and
$$
\oint_{\gamma}d\chi=l(\chi)\in {R}.
$$
Thus we are clearly working in the spaces
$G/H_{\alpha}\times S^1_{\chi}$ for compact
$H_{\alpha}$ and
$G/H_{\alpha}\times {\tilde {R}}_{\chi}$
for non-compact $H_{\alpha}$ for generic
$\alpha$. By looking at the term in the action
quadratic in the $\chi$-field we can determine
the radius of the circle $ S^1_{\chi}$
to be $R_{\chi}=2/r_{\alpha}\sqrt{k\eta_{\alpha}}$.

We can easily check involution using again
the PW property.  The dual model has an
obvious isometry with respect to $\chi$-shifts,
and by applying duality with respect to this
symmetry together with the PW property,
we arrive after a trivial rescaling of variables
to the original theory.

A particularly interesting case corresponds
to the chiral theories $\alpha=0,\pi/2$.  These
theories are self-dual after the first duality
transformation.  For example, for $\alpha=0$,
$J_c=0$ and $\eta_{\alpha}=-t_2$:
$$
{\tilde S}_{\alpha=0}=S_0(g_c)-
{k\over 2\pi}\int {\bar J_c}\del \chi -
{k\over 2\pi}\int t_2\del\chi \delb\chi,
\eqn\ex
$$
and using
the PW property once more we arrive at the
original model. Thus the WZW-model is self-dual
with respect to pure left- or right-abelian
duality.

As a particularly nice example of the previous
considerations we have the duality between
3d-rotating black holes and 3d-charged
black strings
\REF\cosmo{G.T. Horowitz and D.L. Welch:
Phys. Rev. Lett. {\bf 71} (1993), 328.}
[\cosmo].  The 3d-black hole is given
by the gauged WZW-model
$$
\widetilde{SL(2,R)}/SO(1,1)_{\alpha}^Z,
$$
where the notation means that we have to
identify points with integer spacing along
the $SO(1,1)_{\alpha}$-orbits.  $\alpha$ is
related here to the black hole angular momentum,
and we effectively deal with a compact isometry.
$\widetilde{SL(2,R)}$ is the universal covering
of $SL(2,R)$.
Under duality,
$$
\widetilde{SL(2,R)}/SO(1,1)_{\alpha}^Z\rightarrow
\widetilde{SL(2,R)}/SO(1,1)_{\alpha}
\times S^1_{\phi}.
$$
The model on the right is the three-dimensional
charged black string.  For the particular case of
$\alpha=-\pi/4$ we obtain the static three-dimensional
black hole and the uncharged string, which becomes
(at tree level) a simple product of the
two-dimensional black hole and $S^1$.

In conclusion, we have generic topology change
induced by duality even at the semiclassical level.
Quite generally the topological properties of the
dual space depend only on the embedding structure
of the isometry orbits, in particular on its
compactness or non-compactness.  Whether
$\pi_1(M)$ is trivial seems to be, to a large extent
irrelevant.

\chapter{ORDER-DISORDER MAPPING}

Duality transformations map not only the
actions of the $\sigma$-models under consideration
but also provide an explicit dictionary
between the Hilbert spaces, and the physical
operators.  This is the order-disorder
transformation in the context of
Kramers-Wannier duality in Statistical Mechanics,
or the winding-Kaluza-Klein mode mapping
in toroidal compactifications in string theory.
To be more precise, duality in its more general
form is an equality of the form:
$$
\left<O_1...O_n\right>=
\left<{\tilde O}_1...{\tilde O}_n\right>,
\eqn\secfi
$$
for some dual operators ${\tilde O}_i$ to be
determined.  Here we consider local order
operators (in adapted coordinates) of the form
$$
P(\del_z,\del_{\bar z})f(\theta,x^{\alpha})
(z,{\overline z}),
$$
for $f$ some function on the target space-time
$M$ being the analogue of a tachyon operator.
For example, in group manifolds $M=G$,
we may diagonalize the action of $H_{\alpha}$
for the basis functions
$$
f_{p_L,p_R}(g)=f_{p_L,p_R}(e^{\theta_L H}
g_ce^{-\theta_R H})=
e^{i(p_L\theta_L-p_R\theta_R)}f_{p_L,p_R}(g_c).
$$
Since $g_c$ is a spectator field in duality
transformations, we can consider operators
of the form
$$
V_p=e^{ip\theta},
\eqn\fii
$$
and in general,
$$
{\cal O}=P(\del_z,\del_{{\overline z}})
e^{ip\theta}.
\eqn\fii
$$
Thus, in establishing \secfi\ it suffices
to concentrate on tachyon operators.

For non-compact $H_{\alpha}$ $p\in {R}$ (or
more generally $p\in {C}$) while in the compact
case $p\in {Z}$ (assuming that $\theta$
is identified modulo $2\pi$).  It is easy
to perform the duality transformation in the
functional integral (here we consider only
tree-level duality):
$$
\langle V_{p_1}(z_1)\ldots V_{p_n}(z_n)\rangle=
\int D\theta D x^{\alpha}
V_{p_1}(z_1)\ldots V_{p_n}(z_n)e^{-S(\theta,x^{\alpha})}.
$$
First we fix a point on the surface $\Sigma$,
say $P=\infty$ and write,
$$
\prod_i V_{p_i}=\prod_je^{ip_j(\theta(z_j)
-\theta(\infty))}\prod_je^{i p_j\theta(\infty)}
$$
$$
=\prod_je^{ip_j\int_{\infty}^{z_j}d\theta}
e^{i \sum_j p_j\theta(\infty)}.
$$
As usual we may fix the translational symmetry
zero mode in $\theta$ by the requirement of
charge conservation $\sum_jp_j=0$, and the integrals
$\int_{\infty}^{z_j}d\theta=\int_{\gamma_j}d\theta$
go along arbitrary cuts drawn on $\Sigma$.
Now we can use a first order formalism.  We
can further transform the operators to the
form
$$
\int_{\infty}^z d\theta={1\over 2\pi}
\oint_{\gamma_z}d\theta \alpha=
{1\over 2\pi}\int_{\Sigma}d\alpha \wedge
d\theta
$$
for $\alpha$ a function with $2\pi$ jumps on
$\gamma_z$, ie it is an ``angular variable"
centered at $z$.  Now we may gauge the model to
define the following correlator on the
$d+1$-dimensional theory:
$$
\langle \prod_j V_{p_j}\rangle=
\int {D\theta Dx^{\alpha} DA
D\chi\over |H_{\alpha}|}
e^{-S_{eff}(p_j,A,\theta,x^{\alpha},\chi)}
$$
$$
S_{eff}=S_{d+1}(\theta,x_{\alpha},A,\chi)+
{i\over 2\pi}\sum_j p_j \int (d\theta +A)
\wedge d\alpha_j.
$$
Where $|H_{\alpha}|$ is the volume of the gauge
group generated by the isometry.
Thus the dual operators in the $d+1$-dimensional
sense are given by the non-local expressions:
$$
W_p^{(d+1)}(z)=e^{ip\int_{\infty}^z(d\theta+A)}
=e^{ip\int_{\gamma_z}D\theta}.
$$
We can also work out the $d$-dimensional gauge
invariant form by integrating out A. Upon gauge
fixing,
$$
S_{eff}\rightarrow S_{d+1}
(\theta=0,x_{\alpha},A,\phi)
+{1\over\pi}\sum_j p_j\int(A\delb \alpha_j
-{\overline A}\del\alpha_j),
$$
and after integrating over $A$ we obtain a frustrated
partition function
$$
\langle \prod_j V_{p_j}\rangle=
\int Dx^{\alpha} D\phi
e^{-{\tilde S}_d({\phi}^*,x_{\alpha})},
$$
where ${\phi}^*$ is the frustrated field
$$
d{\phi}^*=d\phi+\sum_j 2 p_jd\alpha_j,\qquad
\oint_{|z-z_j|=\epsilon}d{\phi}^*=4\pi p_j.
$$
Hence the dual operators to the $V_p$'s
are genuine winding operators for compact
$H_{\alpha}$ with charge $2p$.  For non-compact
$H_{\alpha}$ we have vortex lines with real charge
$2p$.  This is expected from the study of
the higher genus partition functions.

Recalling the explicit form of the $\phi$-sector
in the dual model,
$$
{\tilde S}_{\phi}\sim {1\over 2\pi}\int
{1\over \xi^2}\del\phi\delb\phi-
{i\over 4\pi}\int {1\over \xi^2}
j\wedge d\phi,
$$
$$
j=(J_c,{\overline J}_c),
$$
we find it is quadratic in $\phi$, so that we
can still sharpen the form of the dual operators.
It proves convenient to choose $\alpha_j$
as solutions to the pull-back Laplace
equation:
$$
d(\xi^{-2}*d\alpha_j)=0.
$$
Then expanding the frustrated action
$$
\eqalign{{\tilde S}_{\phi^*}&={\tilde S}_{\phi}+
{\tilde S}_{\sum_j 2p_j\alpha_j}\cr
&={\tilde S}_{\phi}+\sum_{ij}p_ip_j\int_{\gamma_i}
\xi^{-2}*d\alpha_j-\sum_j{ip_j\over 2\pi}
\int d\alpha_j\wedge j \xi^{-2}.\cr}
$$
Since ${\tilde S}_{\phi}$ is at most quadratic,
and the $\alpha_j$'s are solutions to
the Laplace equation, we can write
$$
\langle \prod_j V_{p_j}\rangle_S=
\langle \prod_j W_{p_j}\rangle _{{\tilde S}},
$$
for
$$
W_p=e^{p\int_{\infty}^z\xi^{-2}*d\phi}.
$$
Note that for constant $\xi^2$, $\phi$ admits
a holomorphic decomposition
$\phi(z,{\overline z})=\phi(z)+{\bar\phi({\overline z})}$,
and we may integrate the operator to
$$
W_p=e^{-{i p\over \xi^2}(\phi(z)-{\bar \phi({\overline z})})},
$$
the usual flat winding mode.  For non-constant
$\xi^2$ this is not possible in general, the operator
remains string-like and the dual operators
take the form
$$
P({1\over ip}\del_z,{1\over ip}\del_{{\overline z}})
e^{-ip\int^z(dz\del_z\phi-d{\overline z}
\del_{{\overline z}}\phi)/\xi^2}.
$$
For example, the dual of $\del\theta\delb\theta
e^{ip\theta}$ is given by
$$
{1\over ip}\del_z\del_{{\overline z}}{\phi\over\xi^2}
e^{-ip\int^z(dz\del_z\phi-d{\overline z}
\del_{{\overline z}}\phi)/\xi^2}.
$$
Two conclusions can be drawn from this analysis:

1).  The semiclassical algorithm allows for an explicit
parametrization of dual operators also for curved
backgrounds, where an exact Conformal Field Theory
construction is not known.  We find in general
non-local operators which cannot be made local
by holomorphic factorization.

2).  The physics of the dual model depends
on the explicit operator mapping.  There is
always a translation dictionary involved.  For
example, for a non-compact isometry group
we have to quantize the dual model as a vortex
gas rather than counting continuous embeddings
in the target space.

\chapter{REMARKS ON DUALITY AND THE COSMOLOGICAL
CONSTANT}

A striking feature of duality is the fact that the
cosmological constant, defined as the
asymptotic value of the curvature scalar is not
in general invariant under duality.  One is
led to wonder to what extent the cosmological
constant is a string observable.  This fact
noticed in
[\cosmo] implies that the usual definition
of $\Lambda$ from the low-energy effective
action is not satisfactory.  Even at large
distances, if duality is not broken there is
a symmetry between local (momentum) modes
and non-local (winding) modes.  The contribution
to the cosmological constant of the massless
sector might be cancelled by the tower of
massive states always present in String Theory
(proposals along these lines using the Atkin-Lehner
symmetry were advanced by G. Moore
\REF\moore{G. Moore: Nucl. Phys. {\bf B293} (1987), 139;
E. Alvarez and M.A.R. Osorio: Z. Phys. {\bf C44}
(1989), 89.}[\moore]).

 We study now the behavior of the scalar curvature
under duality.  If the space-time metric in the
$\sigma$-model takes the form
$$
ds^2=g_{ij}dx^idx^j\qquad
i,j =0,1,2,...,d-1,
\eqn\cosmoi
$$
where $x^0$ is adapted to the isometry
$k=\partial /\partial x^0$, \cosmoi\
can be written as
$$
\eqalign{
ds^2&=(e^0)^2+(g_{\alpha\beta}-{k_{\alpha}k_{\beta}
\over k^2})dx^{\alpha}dx^{\beta}\cr
e^0&=kdx^0+{k_{\alpha}\over k}dx^{\alpha}\cr
k^2&=k_ik^i=g_{00} \qquad k_{\alpha}=g_{0\alpha}.\cr}
\eqn\cosmoii
$$
Buscher's transformation leads to a dual
metric
$$
\eqalign{d{\tilde s}^2&=({\tilde e}^0)^2+(g_{\alpha\beta}-{k_{\alpha}k_{\beta}
\over k^2})dx^{\alpha}dx^{\beta}\cr
{\tilde e^0}&={1\over k}(d{\tilde x}^0
+v_{\alpha}dx^{\alpha})\cr},
\eqn\cosmoiii
$$
${\tilde x}^0$ being the Lagrange multiplier
and $v$ is defined as in section 2 by
$k^lH_{lij}=-\partial_{[i}
v_{j ]}, H=db$.  The dual scalar curvature
following from \cosmoiii\ is
$$
\eqalign{{\tilde R}&=R -{4\over k^2}g^{\alpha\beta}
\partial_{\alpha} k
\partial_{\beta} k +{4\over k}\Delta^{d-1}_q k+\cr
&{1\over k^2}H_{0\alpha\beta}H^{0\alpha\beta}-{k^2\over 4}
F_{\alpha\beta}F^{\alpha\beta},\cr}
\eqn\cosmoiv
$$
where $\Delta^{d-1}_q$ is the $(d-1)$-dimensional
Laplacian for the metric
$g_{\alpha\beta}^q=g_{\alpha\beta}-
{k_{\alpha}k_{\beta}\over k^2}$, and
$F_{\alpha\beta}=\partial_{\alpha} A_{\beta}
 -\partial_{\beta} A_{\alpha}$
with $A_{\alpha}=k_{\alpha}/k^2$.  \cosmoiv\ can be
rewritten as
$$
{\tilde R}=R +4\Delta \log k +
{1\over k^2}H_{0\alpha\beta}H^{0\alpha\beta} -{k^2\over 4}
F_{\alpha\beta}F^{\alpha\beta}.
\eqn\cosmov
$$
{}From \cosmov\ we see that the only way to
``flatten'' negative curvature is by having
torsion in the initial space-time.  Otherwise
the dual of an asymptotically negatively curved space
time is a space of the same type.  Positive
curvature seems however easier to flatten in
view of \cosmov.  We can also construct
the dual torsion
$$
\eqalign{{\tilde H}_{0\alpha\beta}&=-{1\over 2} F_{\alpha\beta}\cr
{\tilde H}_{\alpha\beta\rho}&=H_{\alpha\beta\rho}-
{3\over k^2}H_{0[\alpha\beta}k_{\rho]}-
{3\over 2}F_{[\alpha\beta}v_{\rho]}.\cr}
\eqn\cosmovi
$$
Since
$$
\sqrt{g}=k^2\sqrt{{\tilde g}},
\eqn\cosmovii
$$
and the modulus of $k$ can be expressed
in terms of the dilaton transformation properties,
$$
{\tilde \phi}=\phi -2\log k,
\eqn\cosmoviii
$$
we obtain
$$
{\tilde R}+e^{\phi-{\tilde \phi}}
{\tilde H}^2_{0\alpha\beta} +2 \Delta {\tilde \phi}=
R+e^{{\tilde \phi}-\phi}H_{0\alpha\beta}^2+2\Delta \phi,
\eqn\cosmoix
$$
which can be used to show the duality invariance
of the string effective action to leading order
in $\alpha '$, as first noticed in [\buscher].

The change of the cosmological constant under
duality is not only peculiar to three-dimensions
[\cosmo] but rather generic.  This raises the
physical question of whether in the context
of String Theory the value of the cosmological
constant can be inferred from the asymptotic
(long distance) behavior of the Ricci tensor.
If duality is not broken, the answer
seems to be in the negative, and
it makes the issue of what is the correct
meaning of the cosmological constant in
String Theory yet more misterious.

\chapter{EXAMPLES OF NON-ABELIAN DUALITY}

\section{Generalities and $SU(2)$ Duality}

In this section we would like to explore
some examples and properties of the non-abelian
generalization of duality proposed in
[\queossa].  The philosophy in [\queossa]
is to generalize the gauging procedure in
[\rocver] to manifolds with non-abelian
isometry groups.  This method leads to
an apparent equivalence between two
$\sigma$-models with vastly different topologies
and geometries.  There are some fundamental
differences between abelian and non-abelian
duality transformations which are worth pointing
out.

{\bf 1}.  In the context of Statistical Mechanics,
duality transformations are applied to models defined
on a lattice $L$ with physical variables taking
values on some abelian group $G$.  The duality
transformation takes us from the triplet
$(L,G,S[g])$, where $S[g]$ is the action depending
on some coupling constants labelled collectively
by $g$ to a model $(L^*,G^*,S^*[g^*])$ on the dual
lattice $L^*$ with variables taking values on the
dual group $G^*$ and with some well-defined action
$S^*[g^*]$.  For abelian groups, $G^*$ is the
representation ring, itself a group, and when
we apply the duality transformation once again
we obtain the original model.  As soon as
the group is non-abelian the previous construction
breaks down because the representation ring
of $G$ is not a group [1].  In particular the
non-abelian duality transformations
cannot be performed again to
obtain the model we started with.

{\bf 2}.  In the continuum, and for abelian
duality, the number of abelian symmetries in
the original and the dual models is the same.
The dual abelian isometries correspond to
constant shifts of the associated Lagrange
multipliers (the non-abelian part of the original
isometry group is probably realized non-locally).
In the non-abelian case we systematically
lose symmetries.  The non-abelian isometry
group of the dual space is always smaller, and
the original isometry group is not realized
locally in the dual theory.  It is a difficult
problem to find the original theory if we are
only given its non-abelian dual.

For the above two reasons it is somewhat of
a misnomer to call non-abelian duality a duality
transformation.  We will nevertheless follow
the nomenclature in the literature.  We should
also mention that some of the most interesting
examples considered below are not conformally
invariant field theories, although one could
imagine in principle their embedding in conformally
invariant $(2,0)$ or $(2,2)$ supersymmetric
$\sigma$-models.  We have not attempted to
carry this out.

Following the outline in [\queossa]  we begin
with the standard $\sigma$-model on a group
manifold $G$.
$$
S(g)=-{k\over 2\pi}\int Tr (\gdg \gdbg).
\eqn\nadi
$$
The action is invariant under the left- and
right-action of elements of $G$, $g\rightarrow
h_1gh_2^{-1}$. $k$ in \nadi\ needs not
be an integer since we are not including
the WZW term.
Since we want to maintain as far as
possible the background independence of our
manipulations, we perform non-abelian duality
with respect to some sugbroup $H$ of $G$ acting
without fixed points.  For instance we
can take a subgroup $H$ acting by left-multiplication.
Gauging $H$ and introducing Lagrange multipliers
in its adjoint representation, we obtain:
$$
\eqalign{S(g,A,\chi)&=-{k\over 2\pi}\int Tr( \gdg \gdbg
+A \delb g g^{-1} +{\overline A}\del g g^{-1}+\cr
&{\overline A}A +\chi (\del {\overline A}-\delb A
+[A,{\overline A}])).\cr}
\eqn\nadii
$$
\nadii\ is invariant under the gauge transformations
$$
g\rightarrow hg\qquad A\rightarrow h(A+d)h^{-1}
\qquad \chi\rightarrow h\chi h^{-1}.
\eqn\nadiii
$$
To explore non-abelian duality we choose the largest
non-abelian subgroup of $G$
acting without fixed points,
namely $G$ itself
acting on the left.  Rather than solving for $A$
in \nadii\ and then gauge fixing, we first make
the change of variables $A\rightarrow g^{-1}(A+d)g,
\chi\rightarrow g^{-1}\chi g$.  Then $g$ disappears
from the action.  This is equivalent to choosing
the unitary gauge $g=1$.  The gauge fixed action
becomes
$$
S(A,\chi )=-{k\over 2\pi}\int Tr({\overline A}A+
\chi F(A)).
\eqn\nadiv
$$
Using the equations of motion for $A$,
$$
{\overline A}+\delb \chi +[{\overline A},\chi ]=0,
\eqn\nadv
$$
\nadiv\ becomes
$$
S(A,\chi )={k\over 2\pi}\int Tr ({\overline A}\del \chi ).
\eqn\nadvi
$$

Solving ${\overline A}$ in terms of the Lagrange
multipliers from \nadv\ leads to the dual action.
Note that for compact groups \nadv\ can always
be solved without singularities.

The obvious example to consider is $G=SU(2)$. Take
as generators $T_i=\sigma_i/i\sqrt{2}$.  ${\overline A}$
satisfies:
$$
(\delta^{ij}+\sqrt{2} \epsilon^{ijk} {\chi}^k){\overline A}^j
=-\delb {\chi}^j.
\eqn\nadvii
$$
Solving for ${\overline A}$ and substituting in
\nadvi\ leads to
$$
{\tilde S}(\chi )={k\over 4\pi}\int \left[
{1\over 1+{\chi}^2}(\delta_{ij}+{\chi}_i{\chi}_j)-
{1\over 1+{\chi}^2}\epsilon_{ijk}{\chi}_k\right]
\del {\chi}^i \delb {\chi}^j.
\eqn\nadviii
$$
In \nadviii\ ${\chi}^2=\delta_{ij}{\chi}^i{\chi}^j$,
and we have
absorbed a factor of $\sqrt{2}$ in $\chi$.  The dual
metric is therefore,
$$
d{\tilde s}^2={1\over 1+{\chi}^2}
(\delta_{ij}+{\chi}_i{\chi}_j)d{\chi}^id{\chi}^j.
\eqn\nadix
$$
Using polar coordinates,
$$
d{\tilde s}^2=dr^2+{r^2\over 1+r^2}(d\theta ^2+\sin^2\theta
d\phi ^2).
\eqn\nadx
$$
Asymptotically as $r\rightarrow \infty$ this is the
standard metric on ${R}\times S^2$, and for
$r\rightarrow 0$ the space looks like ${R}^3$.  The
scalar curvature for \nadx\
$$
R=2{(9+3r^2+r^4)\over (1+r^2)^2},
\eqn\nadxi
$$
is free of singularities as expected because
the group acts without fixed points.  Hence, at
tree level, the non-abelian dual of $G=SU(2)$
with respect to its left-action is the Lie
algebra of $SU(2)$ but with a metric
reminiscent of Witten's $SL(2,{R})/U(1)$ Euclidean
black hole metric
\REF\blackh{E. Witten:
Phys. Rev. {\bf D44} (1991), 314.}
[\blackh].  According to [\queossa]
the dilaton changes into
$$
{\tilde \phi}=\phi-\log(1+{\chi}^2).
$$
Notice that ${\tilde S}(\chi )$ contains less symmetry
than the original model.  \nadi\ is invariant
under $g\rightarrow h_1gh_2^{-1}$.  Since
the dual variables are the gauge invariant
combinations $g^{-1}\chi g$, we are left with
the action of $SU(2)$ on the right.  We can proceed
to perform the duality transformation with
respect to this residual group.  Now $\chi =0$
is a fixed point of the isometry group
and the resulting space will be singular.  We
gauge the symmetry by replacing derivatives
by covariant derivatives $D\chi =\del \chi +[A,\chi ]$,
and then we add the new Lagrange multiplier
$\lambda$ to obtain
\def\ts{{\tilde S}}
$$
\ts (\chi ,A,\lambda)={k\over 4\pi}
\int\left(-{Tr (D\chi {\overline D}\chi )\over 1+\rho^2}
-{1\over 1+\rho^2}\epsilon_{ijk}{\chi}_k
D{\chi}_i{\overline D}{\chi}_j+{\del\rho \delb\rho\over
1+\rho^2}+ Tr\lambda F(A)\right),
\eqn\nadxii
$$
with
$$
\rho = \sqrt{{\chi}_i{\chi}_i}.
$$
Before integrating the gauge field, we fix the
non-abelian part of the gauge by choosing the
unitary-like gauge
$$
\chi={1\over i\sqrt{2}}\pmatrix{\rho&0\cr
0&-\rho\cr}.
\eqn\nadxiii
$$
We fix the residual $U(1)$ gauge invariance as
follows: Since $\lambda=\lambda_a\sigma^a/i\sqrt{2}$,
$$
\lambda={1\over i\sqrt{2}}\pmatrix{\lambda_3&
\lambda_1-i\lambda_2\cr
\lambda_1+i\lambda_2&-\lambda_3\cr},
\eqn\xiv
$$
write
$$
\lambda_1=r\cos\theta\qquad
\lambda_2=r\sin\theta,
\eqn\nadxv
$$
and choose the gauge
$$
\theta=0
\eqn\nadxvi
$$
This fixes the gauge completely and the new
dual variables are $(\rho,r,\lambda_3)$.
Eliminating $A$ and after some manipulations
we obtain
$$
{\tilde \ts}={k\over 4\pi}\int (
\del\rho\delb\rho+{1+\rho^2\over 2\rho^2}
\del r\delb r+
\left({1+\rho^2\over 2r \rho^2}\lambda_3+
{\rho\over \sqrt{2} r}\right)
$$
$$
\left(\del r \delb \lambda_3 +
\del \lambda_3 \delb r\right)
+\left({1+\rho^2\over 2r^2 \rho^2}\lambda_3^2
+{\rho^2\over r^2}+{\sqrt{2}\rho\over r^2}
\lambda_3\right)\del\lambda_3
\delb\lambda_3).
\eqn\nadxvii
$$
The new dilaton is
$$
{\tilde {\tilde \phi}}=\phi - \log{{{\rho}^2 r^2
\over 2}}.
\eqn\nadxviii
$$
The double dual has curvature singularities
at $\rho=0$ and $r=0$.  The Lagrangian
\nadxvii\ has no trace of any of the original
$SU(2)_L\times SU(2)_R$ symmetries or
of the $SU(2)$ symmetry of the dual Lagrangian.
There are no symmetries left at all.  It is
quite unclear also how to go back from the
dual to the original action in contrast with the
abelian case.

\section{The Dual of $SL(2,{R})_L$}

A more interesting example from the space-time
point of view is as above but with
$SU(2)$ replaced by $SL(2,{R})$.  The result
is simply obtained by analytically
continuing the $SU(2)$ formulas according
to the rules ${\chi}_1\rightarrow i{\chi}_1$,
${\chi}_2\rightarrow {\chi}_2$, ${\chi}_3
\rightarrow i{\chi}_3$
(plus a change $k\rightarrow -k$ to have
signature $(-,+,+)$).  We obtain the
dual Lagrangian
$$
{\tilde S}(\chi )={k\over 4\pi}\int \left[
{1\over 1-{\chi}^2}(\eta_{ij}-{\chi}_i{\chi}_j)-
{1\over 1-{\chi}^2}\epsilon_{ijk}{\chi}_k\right]
\del {\chi}^i \delb {\chi}^j,
\eqn\nadxix
$$
$$
{\chi}^2=\eta^{ij}{\chi}_i{\chi}_j\qquad
\eta={\rm diag}(1,-1,1).
\eqn\nadxx
$$
Using the analogue of polar coordinates
we can distinguish two regions:
${\chi}^2>0$ and ${\chi}^2<0$.  The dual metric takes
the form
$$
\eqalign{{\chi}^2>0,\qquad d{\tilde s}_I^2&={1\over 4\rho}d\rho^2
-{\rho\over 1-\rho}d\eta^2+{\rho\over 1-\rho}
{\rm cosh}^2\eta d\phi^2\cr
&\rho ={\chi}^2;\qquad \eta\in {R};
\qquad \phi\in[0,2\pi]\cr
{\chi}^2<0\qquad d{\tilde s}_{II}^2&=-{1\over 4\rho}d\rho^2+
{\rho\over 1+\rho}d\eta^2+{\rho\over 1+\rho}
{\rm sinh}^2\eta d\phi^2\cr
&\rho=-{\chi}^2 , \cr}
\eqn\nadxxi
$$
with in region I
$$
\eqalign{ \chi_1&=\rho^{1/2}\cos\phi\cosh\eta\cr
\chi_3&=\rho^{1/2}\sin\phi\cosh\eta\cr
\chi_2&=\rho^{1/2}\sinh\eta ,\cr}
$$
and similarly in region II with minor changes.
The scalar curvatures are
$$
R_I=-2{(\rho^2-3\rho+9)\over (\rho-1)^2}
\qquad
R_{II}=-2{(\rho^2+3\rho+9)\over (\rho+1)^2}.
\eqn\nadxxiii
$$
$R_I$ has a singularity at $\rho=1$.  This is
not due to fixed points but rather to the
non-compactness of the group (infrared problem).
The dual space has interesting global properties.
It is an anti-DeSitter (AdS) space-time.
$\rho=0$ is a coordinate singularity where the metric
goes from $(-,+,+)$ in region II to
$(+,-,+)$ in region I, and $(+,+,-)$ beyond
the singularity.  Thus $\rho=0$ is likely
to be a horizon.  To study the causal
structure we can ignore the $\phi$-dependence.
Each point $(\rho,\eta)$ is a circle of
radius $(\rho/(1-\rho))^{1/2}\cosh{\eta}$
in region I and of radius
$(\rho/(1+\rho))^{1/2}|\sinh{\eta}|$
in region II.  Making $\rho\rightarrow -\rho$
in region II we cover the space-time with
$$
d{\tilde s}^2=-{\rho\over 1-\rho}d\eta^2
+{1\over 4\rho^2} d\rho^2.
\eqn\nadxxiv
$$
The $\rho=0$ coordinate singularity can
be eliminated by going to ``tortoise''
coordinates
$$
\rho^*={1\over 2}\log{|{1-\sqrt{1-\rho}\over
1+\sqrt{1-\rho}}|}+\sqrt{1-\rho};\qquad \rho<1,
\eqn\nadxxv
$$
so that
$$
d{\tilde s}^2=
-{\rho\over 1-\rho}d\eta^2+
{\rho\over 1-\rho}d{{\rho}^*}^2.
\eqn\nadxxvi
$$
Region II in these coordinates is covered
by $-\infty<\rho^*<\infty$.  Going to
null coordinates $u=\rho^*-\eta,
v=\rho^*+\eta$ , and using Kruskal-like
coordinates,
$$
{\overline u}=e^u\qquad {\overline v}=e^v,
\eqn\xxvii
$$
region II is covered with $0<{\overline u}<
\infty, 0<{\overline v}<\infty$.  In terms
of these coordinates we can extend the manifold
to a larger manifold
\def\mtil{{\tilde M}}
$\mtil$
$$
d{\tilde s}^2={\rho\over 1-\rho}
e^{-2\rho^*}d{\overline u}d{\overline v}
\qquad -\infty<{\overline u},{\overline v}<
\infty .
\eqn\nadxxviii
$$
The extended region
$-\infty<{\overline u},{\overline v}<0$
is isomorphic to region I and the transition
takes place analytically.  In Kruskal coordinates
the horizon is at ${\overline u}=0$,
${\overline v}=0$.  The Penrose diagram
for $\mtil$ appears in fig. 1.


\midinsert
\epsfysize=3in
\centerline{\epsffile{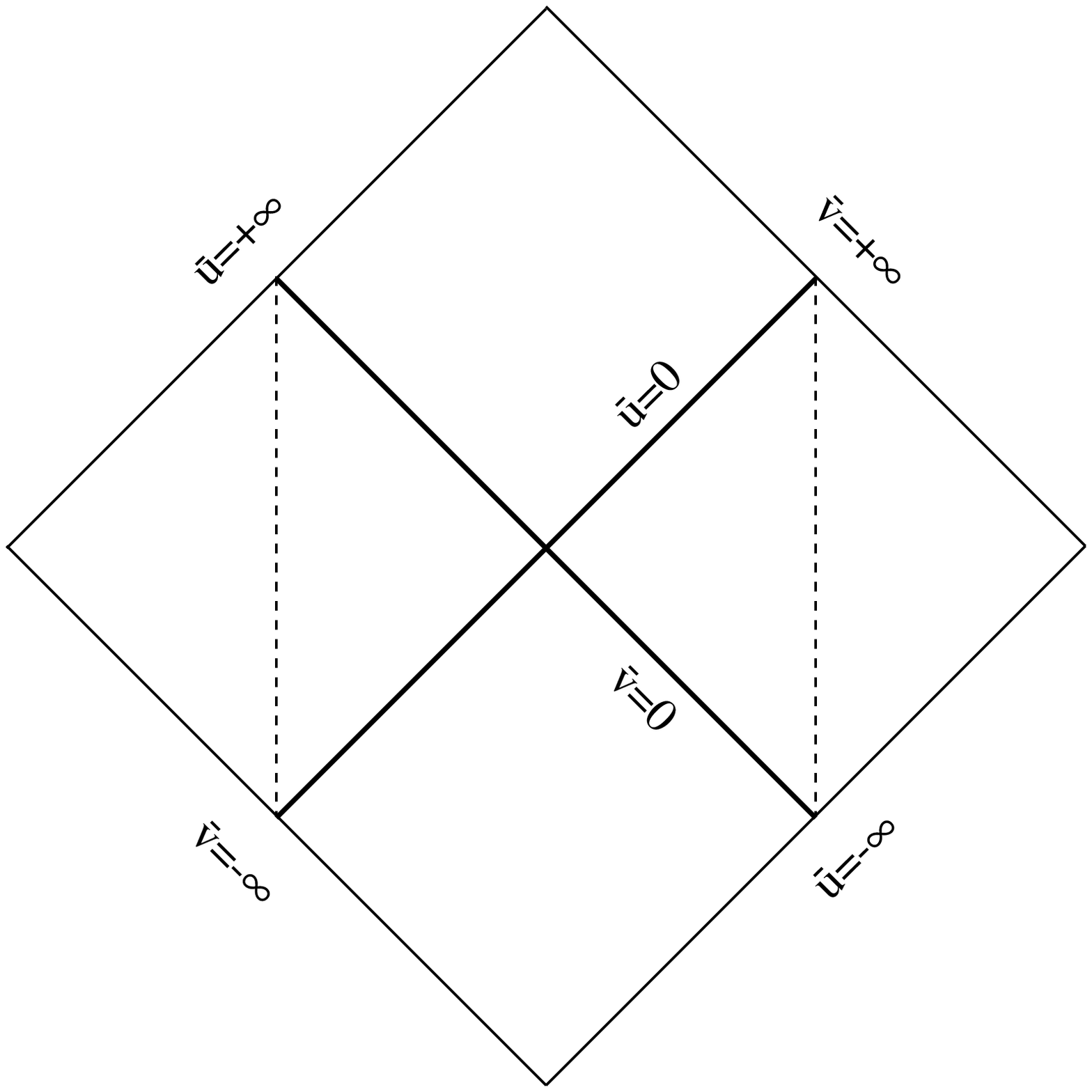}}
\caption{Fig.1: Penrose diagram for ${\tilde M}$.}
\endinsert

The Penrose diagram describes a black-hole
space-time with a time-like singularity.
Asymptotically ${\tilde M}$ has negative
scalar curvature with metric
$ds^2=-d\rho^2+d\eta^2+\sinh^2\eta d\phi^2$.
We have thus found a black-hole space-time
by dualizing AdS space with respect to the
non-abelian symmetry group $SL(2,{R})_L$.
This space-time is different from the
three-dimensional space-time found in
\REF\threedbh{M. Ba\~nados, M. Henneaux,
C. Teitelboim and J. Zanelli: {\it Geometry
of the 2+1 Black Hole}.
To appear in Phys. Rev. D.}[\threedbh]
which arises from identifying points
in AdS space by a discrete subgroup of
$SO(2,2)$.  A time-like singularity
appears there, but it is not a curvature singularity.
It is a causal singularity beyond which
closed time-like curves appear.

In the dual model we still have an
$SL(2,{R})$-isometry group, and we can
again dualize with respect to it.  The
formulae are similar to those obtained
for the $SU(2)$ case after analytic
continuation.  We will not spell out
the details here.

An important remark is that even though
the space-time obtained is potentially
interesting in the subject of string black
holes, we have obtained the result starting
with a theory which is not conformally
invariant.  It is likely that conformal
invariance can be gained by embedding the
model in a $(2,0)$ supersymmetric theory,
but we have not yet explored this possibility.

An example where one can start directly with
a conformally invariant field theory is
the $SL(2,{R})$-WZW theory if we
perform duality
with respect to the vector action of
$SL(2,{R})$.
This model leads to a rather singular space-time
and presumably the correct way to think about
the dual theory is without the elimination
of the gauge field and using BRST techniques.
In this model the gauge fixing is rather
delicate because we have to distinguish
the elliptic, parabolic and hyperbolic
regions of $SL(2,{R})$.
We have not found
a nice physical interpretation of the dual
space-time as a black-hole or any other
simple space-time, and will just present
an outline of the results in the appendix.

\section{Non-Abelian Frustrations}

To conclude this chapter we would like
to address some of the problems in the
determination of the global properties
of non-abelian dual spaces and the current
difficulties one finds in trying to obtain
the operator mapping.  In particular
it should also be interesting to determine
the form of this transfomation for higher
genus Riemann surfaces.  Given that duality
transformations are versions of Kramers-Wannier
duality, it seems at first sight paradoxical
that no use of non-abelian duality has been
made in Statistical Mechanics.  The reason
is simple as mentioned at the beginning of
the chapter.  For non-abelian groups the
set of representations does not form a group.
To be more precise, duality always proceeds
by first transforming to a first order
formalism (using gauge fields) with constraints.
When the constraints are solved we get back
the original model.  In the abelian case the
constraints take the form
$$
\prod_z\delta(dA)\prod_i\delta(\oint_{a_i}A)
 \delta(\oint_{b_i}A),
\eqn\nadxxix
$$
where $a,b$ are the non-trivial homology cycles
of a genus-g Riemann surface.  In the non-abelian
case similar gauge invariant constraints are
expressed in non-local form in terms of
Wilson loops (we think of hte theory as defined
on a lattice)
$$
\prod_{{\rm plaquettes}}
\delta(W_{{\rm plaquettes}})
\prod_i\delta(W_{a_i})\delta(W_{b_i}),
\eqn\nadxxx
$$
where the $a,b$'s are now understood as generators
of the homotopy of the surface, and
$$
W_{\gamma}=Pe^{\int_{\gamma}A}.
\eqn\nadxxxi
$$
Delta-functions in \nadxxix\ can be expanded
in Fourier modes and we obtain local terms
in the action for the Lagrange multipliers.  However
for the non-abelian delta-functions
this is not the case.  For instance, in the
case of $SU(2)$, we use the character
expansion to obtain
$$
\delta(g)=\sum_l\sum_{m,n=-l}^l
t^l_{mn}(g){\overline t^l_{mn}(1)},
\eqn\nadxxxii
$$
where
$$
t^l_{mn}(g)=e^{i(m\phi+n\psi)}
P_{mn}(\cos\theta),
$$
in terms of Euler angles.  We see that
only the $U(1)$ embeddings parametrized
by the angles $\phi,\psi$ exponentiate in local
form ($g$ represents a local product
of link variables around a plaquette).
Because of the infinite product of Legendre
polynomials we do not know how to perform
the link integral (gauge field integral)
and obtain a local model in the fields
$l(z,{\overline z}),n(z,{\overline z}),
m(z,{\overline z})$ defined on the dual
lattice.

In the genus zero case and in the continuum
we can circumvent the problems by using
covariant Lagrange multipliers as we did
previously.  This allows us to project
into the space of flat connections, however
the problem reappears with the homotopy
Wilson loops in the higher genus case,
or even at tree level in the presence
of operator insertions.  These are serious
problems in order to state non-abelian
duality as a full-fledged string symmetry.
In the particular case of genus one one can
go a little further by using the fact that
the fundamental group is abelian and flat
connections can therefore be gauge
transformed into the Cartan subalgebra of the
group, thus allowing a treatment similar
to the abelian case.  When one tries to carry
in detail the computations for genus
higher than zero, one finds that the basic
difficulty is the fact that the Hodge decomposition
and the splitting of the flat connection part
of the auxiliary gauge field only coincide for
the abelian case.  In some cases however
one can study in some detail the global
topological properties of the dual model directly.

The same type of problems appear when we try
to carry out in detail the operator mapping
between a given theory and its non-abelian
dual.  For instance, for a $\sigma$-model
on a group manifold $G$, we could consider
correlators of the form
$$
\langle \prod_jg(z_j,{\overline z}_j)\rangle,
$$
which can be expressed in the first order
formalism by choosing a point $P=\infty$
and a system of cuts $\gamma_j$ going from it
to $z_j$.  Fixing $g(\infty)=1$ we can write
$$
\langle \prod_jg(z_j,{\overline z}_j)\rangle=
\langle \prod_j W_{\gamma_j}(g^{-1}dg)\rangle,
$$
for
$$
W_{\gamma_j}(g^{-1}dg)=Pe^{\int_{\gamma_j}
g^{-1}dg},
$$
and in the gauge model we have
$$
\langle \prod_jg(z_j,{\overline z}_j)\rangle=
\int{Dg DA D\chi\over |G|}e^{-S(g,A,\chi)}
\prod_j W_{\gamma_j},
$$
with
$$
W_{\gamma_j}=Pe^{\int_{\gamma_j}g^{-1}Dg}
$$
$$
Dg=dg+Ag,
$$
in the case that the isometries act by
left multiplication.  As usual we can take
these operators as duals to the $g(z_i)$
but still we cannot perform the gauge field
integration and express them in terms of the
Lagrange multipliers due to the presence of
the path ordering prescription.

It is quite clear that more work is needed
to elucidate the complete structure of the
non-abelian duality transformation.

{\bf Acknowledgements}. We would like to
thank X. De La Ossa, M.A.R. Osorio, F. Quevedo
and M.A. V\'azquez-Mozo for very useful
discussions.
E. Alvarez is grateful to P.
Minkowski, P. Hajicek and
other members of the ITP in Bern Universit\"at
for their hospitality while
most of this work was done.  His work
was supported in part by the Tomalla Foundation
(Switzerland) and by CICyT (Spain),
contract number AEN/93/673.
J.L.F. Barb\'on and Y. Lozano
would like  to thank
the Theory Division at CERN for
its hospitality
while part of this
work was done. J.L.F.B.
was supported in part by a
Fellowship from M.E.C. (Spain).
Y.L. was supported
in part by a
Fellowship from
Comunidad de Madrid ( Spain).
\endpage

\appendix

The vectorially gauged action reads:
$$
\eqalign{S(g,A,\chi )&=S(g)+{k\over 2\pi}\int
Tr(A{\bar \partial}g g^{-1}-{\bar A}
g^{-1}\partial g-g^{-1}A g{\bar A}+\cr
&A {\bar A}+\chi (\partial {\bar A}-
{\bar \partial}A+[A,{\bar A}]))}
\eqn\vecto
$$

\vecto\ is invariant under local gauge
transformations:
$$
g\rightarrow h^{-1}gh; \qquad A\rightarrow
h^{-1}(A+d)h; \qquad \chi\rightarrow
h^{-1}\chi h
\eqn\transfo
$$
with $h\in SL(2,{R})$.
To gauge-fix this action we must choose a
representative for every value of the trace
and then fix the residual $U(1)$-symmetry
on the lagrange multipliers.

Integration over the gauge fields yields the
following result for the elliptic,
hyperbolic and parabolic regions of
$SL(2,{R})$:

Elliptic region:
$$
\eqalign{{\tilde S}&={k\over 2\pi}\int {1\over {\rho}^2
\sin^2\theta}[{\rho}^2\partial\rho
{\bar \partial}\rho-\rho (\chi+{1\over 2}
\sin{2\theta})(\partial\rho {\bar \partial}\chi+
{\bar \partial}\rho \partial \chi)- \cr
&\rho (\chi+
{1\over 2}\sin{2\theta})(\partial\rho
{\bar \partial}\theta+{\bar \partial}\rho
\partial\theta)+({\chi}^2+\chi \sin{2\theta}+
\sin^2\theta)\partial \chi
{\bar \partial}\chi +\cr
&({\chi}^2+\chi \sin{2\theta}+
\sin^2\theta)(\partial
\chi {\bar \partial }\theta+
{\bar \partial}\chi\partial\theta)+
({\chi}^2+\chi \sin{2\theta}+
(1-{\rho}^2)\sin^2\theta)
\partial\theta {\bar \partial}\theta ] \cr
&\theta\in [0,2\pi];\qquad \rho ,\chi \in {R}}
\eqn\elliptic
$$

Hyperbolic region:
$$
\eqalign{{\tilde S}&=\pm {k\over 2\pi}\int {1\over
{\rho}^2 \sinh^2t}[{\rho}^2\partial\rho
{\bar \partial}\rho\mp\rho (\chi+{1\over 2}
\sinh{2t})(\partial\rho {\bar \partial}\chi+
{\bar \partial}\rho \partial \chi )\cr
&\mp\rho (\chi+
{1\over 2}\sinh{2t})(\partial\rho
{\bar \partial}t+{\bar \partial}\rho
\partial t)+({\chi}^2+\chi \sinh{2t}+
\sinh^2t)\partial \chi
{\bar \partial}\chi +\cr
&({\chi}^2+\chi \sinh{2t}+\sinh^2t)(\partial
\chi {\bar \partial }t+
{\bar \partial}\chi\partial t)+
({\chi}^2+\chi \sinh{2 t}+(1\pm{\rho}^2)\sinh^2t)
\partial t {\bar \partial}t] \cr
&t, \rho, \chi\in {R}}
\eqn\hiperb
$$

Parabolic region:
$$
\eqalign{{\tilde S}&={k\over 2\pi}\int
{1\over x^2} [\partial x{\bar \partial}x+
\partial y{\bar \partial}y\pm (\partial x
{\bar \partial}y+\partial y{\bar \partial}x)] \cr
&x,y\in {R}}
\eqn\parab
$$
The new dilatons are:
$$
\eqalign{{\tilde\phi}&=\phi-\log{(8{\rho}^2
\sin^2\theta)}\cr
{\tilde\phi}&=\phi-\log{(8{\rho}^2
\sinh^2t)}\cr
{\tilde\phi}&=\phi-\log{(2x^2)}}
\eqn\dils
$$
The singular points in each region correspond to the
fixed points of the vectorial action.
The dual metric in the parabolic region is
singular, which means that in this region there is
only one propagating degree of freedom. Since the
parabolic region has zero measure in the
whole dual space-time, this loss of degrees of freedom
is possibly irrelevant.

The disconnectedness of the three regions makes unclear
the interpretation of the dual space-time.
One could also choose a different gauge fixing
which does not distinguish between the three different
regions in $SL(2,R)$, however one finds the same problem
later when fixing the residual gauge symmetry
action on the Lagrange multipliers.
A possible way to treat the dual theory could be without
eliminating the gauge-fields, using BRST techniques.

\refout
\end